\documentclass[prd,twocolumn,twoside,preprintnumbers,superscriptaddress,nofootinbib]{revtex4}
\usepackage{amsmath,slashed}
\usepackage{graphicx,graphics,color}
\usepackage{dcolumn}
\usepackage[hyperfootnotes=false]{hyperref}
\usepackage{xspace}

\newcommand{\Order}{\mathcal{O}}

\newcommand{\Lagr}{\mathcal{L}}
\newcommand{\beq}{\begin{equation}}
\newcommand{\eeq}{\end{equation}}

\newcommand{\Ft}{\mathcal{F}t}
\newcommand{\eps}{\varepsilon}

\allowdisplaybreaks[1]

\begin{document}

\preprint{INT-PUB-20-005, PSI-PR-20-02, UZ-TH 05/20}
\title{Beta decays as sensitive probes of lepton flavor universality}

\author{Andreas Crivellin}
\affiliation{Paul Scherrer Institut, CH--5232 Villigen PSI, Switzerland}
\affiliation{Physik-Institut, Universit\"at Z\"urich, Winterthurerstrasse 190, CH--8057 Z\"urich, Switzerland}
\author{Martin Hoferichter}
\affiliation{Albert Einstein Center for Fundamental Physics, Institute for Theoretical Physics, University of Bern, Sidlerstrasse 5, CH--3012 Bern, Switzerland}
\affiliation{Institute for Nuclear Theory, University of Washington, Seattle, WA 98195-1550, USA}

\begin{abstract}
Nuclear $\beta$ decays as well as the decay of the neutron are well-established low-energy probes of physics beyond the Standard Model (SM). In particular, with the axial-vector coupling of the nucleon $g_A$ determined from lattice QCD, the comparison between experiment and SM prediction is commonly used to derive constraints on right-handed currents. Further, in addition to the CKM element $V_{us}$ from kaon decays, $V_{ud}$ from $\beta$ decays is a critical input for the test of CKM unitarity. 
 Here, we point out that the available information on $\beta$ decays can be re-interpreted as a stringent test of lepton flavor universality (LFU). In fact, we find that the ratio of $V_{us}$ from kaon decays over $V_{us}$ from $\beta$ decays (assuming CKM unitarity) is extremely sensitive to LFU violation (LFUV) in $W$--$\mu$--$\nu$ couplings thanks to a CKM enhancement by $(V_{ud}/V_{us})^2\sim 20$. From this perspective, recent hints for the violation of CKM unitarity can be viewed as further evidence for LFUV, fitting into the existing picture exhibited by semi-leptonic $B$ decays and the anomalous magnetic moments of muon and electron. Finally, we comment on the future sensitivity that can be reached with this LFU violating observable and discuss complementary probes of LFU that may reach a similar level of precision, such as $\Gamma(\pi\to\mu\nu)/\Gamma(\pi\to e\nu)$ at the PEN and PiENu experiments or even direct measurements of $W\to\mu\nu$ at an FCC-ee.   
\end{abstract}

\maketitle

{\it Introduction}.---Within the SM of particle physics the masses and mixing angles of quarks have a common origin: their Yukawa couplings to the Higgs boson. In the physical basis with diagonal mass matrices, the misalignment between the up- and down-quark Yukawa couplings is parameterized by the unitary Cabibbo--Kobayashi--Maskawa (CKM) matrix~\cite{Cabibbo:1963yz,Kobayashi:1973fv}. Therefore, the elements of the CKM matrix are fundamental quantities of the SM and their determination is of utmost theoretical and experimental importance~\cite{Tanabashi:2018oca}.

\begin{figure}[t]
	\centering
	\includegraphics[width=0.8\linewidth]{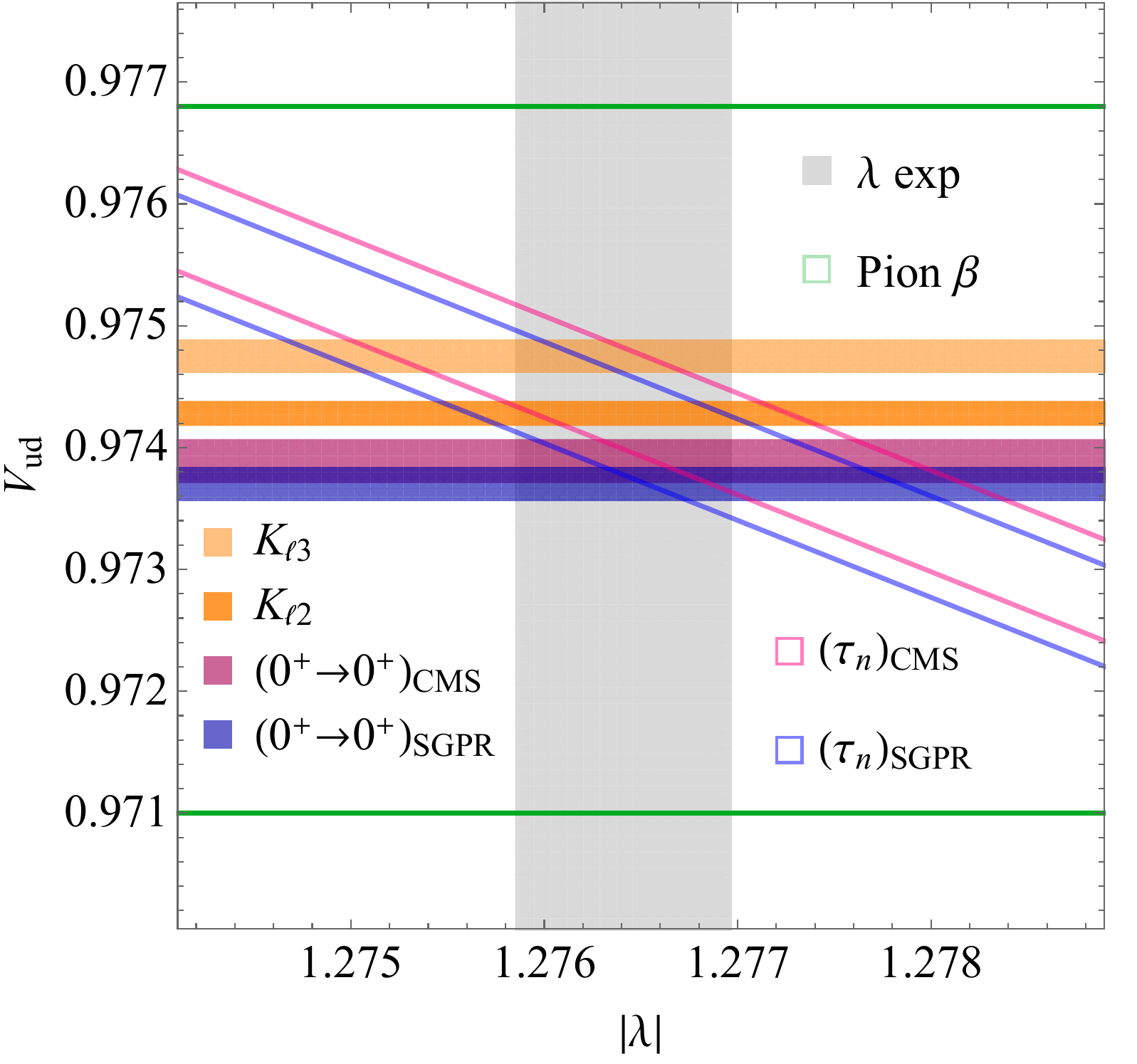}
	\caption{Constraints on $V_{ud}$ and $|\lambda|$ from superallowed $\beta$ decays~\cite{Hardy:2016vhg,Hardy:2018zsb}, neutron life time~\cite{Pattie:2017vsj}, and asymmetry parameter~\cite{Markisch:2018ndu}, for two sets of radiative corrections~\cite{Seng:2018yzq,Czarnecki:2019mwq}. We also include the constraints from pion $\beta$ decay~\cite{Pocanic:2003pf,Czarnecki:2019iwz} and as well as from $V_{us}$ as determined from $K_{\ell 2}$  and $K_{\ell 3}$  decays when assuming CKM unitarity. On this scale, the determination of $g_A$ from lattice QCD~\cite{Chang:2018uxx} does not yet provide a competitive constraint.}
	\label{fig:Vudlambda}
\end{figure}

Superallowed $\beta$ decays---long-lived nuclear $0^+\to 0^+$ transitions---are the primary source of information on the CKM matrix element $V_{ud}$~\cite{Hardy:2014qxa,Hardy:2016vhg,Hardy:2018zsb}. However, additional information can be obtained from neutron decay, whose life time and decay asymmetry parameter together determine the axial-vector coupling of the nucleon $g_A$ as well as $V_{ud}$, with a sensitivity close to, but not yet competitive with, the superallowed nuclear decays~\cite{Czarnecki:2018okw}. The present situation is illustrated in Fig.~\ref{fig:Vudlambda}, based on recent results for the neutron life time
$\tau_n=877.7(7)(^{+0.4}_{-0.2})\,\text{s}$~\cite{Pattie:2017vsj}
and the asymmetry parameter 
$\lambda=g_A/g_V=-1.27641(56)$~\cite{Markisch:2018ndu}.
The values of $V_{ud}$, both from superallowed $\beta$ decays and the neutron lifetime, depend crucially on the applied radiative corrections~\cite{Marciano:2005ec,Seng:2018yzq,Seng:2018qru,Gorchtein:2018fxl,Czarnecki:2019mwq}. Here, we will consider two sets of corrections ``SGPR''~\cite{Seng:2018yzq} and ``CMS''~\cite{Czarnecki:2019mwq} to illustrate the spread. The agreement of the bands from  $0^+\to 0^+$ transitions and neutron decays improves if the value for the neutron life time is moved towards its PDG value $\tau_n=879.4(6)\,\text{s}$~\cite{Tanabashi:2018oca}, highlighting the importance of an accurate $\tau_n$ measurement for the $V_{ud}$ determination. The figure also shows the constraint from pion $\beta$ decay $\pi^\pm \to \pi^0 e^\pm \nu_e$~\cite{Pocanic:2003pf,Czarnecki:2019iwz} as well as the preferred values for $V_{ud}$ deduced from $K_{\ell 2}$ ($K\to \ell\nu$, $\ell=\mu,e$) and $K_{\ell 3}$ ($K\to\pi \ell\nu$) decays under the assumption of CKM unitarity. The observed discrepancy between the latter and the direct determinations of $V_{ud}$ has been interpreted as a possible sign for the (apparent) violation of CKM unitarity and triggered recent interest in potential explanations beyond the SM (BSM)~\cite{Belfatto:2019swo,Cheung:2020vqm}. However, inducing a sizable violation of CKM unitarity is in general difficult due to the strong bounds from flavor-changing neutral currents, such as kaon mixing, see, e.g., Ref.~\cite{Bobeth:2016llm}. 

In addition, the determination of $g_A$ from neutron decay, once compared to calculations in lattice QCD~\cite{Chang:2018uxx}, allows one to constrain the size of right-handed $ud$ currents~\cite{Alioli:2017ces}, albeit not yet at a level competitive with the experimental determination of $g_A$. Still, the comparison shows that both determinations of $g_A$ are compatible, demonstrating that within current uncertainties there is no evidence for right-handed contributions. In particular, their effect does not suffice to remove the tension with CKM unitarity, although a small reduction in significance is possible given that $K_{\ell 2}$ and $K_{\ell 3}$  decays are affected in the opposite way, so that their determinations of $V_{us}$ can be brought into better agreement~\cite{Grossman:2019bzp}.  This situation is reminiscent of previous hints for right-handed currents in semi-leptonic $B$ decays~\cite{Crivellin:2009sd,Buras:2010pz} that disappeared with updated measurements and theory predictions, to the effect that currently an explanation of the discrepancies between the inclusive and exclusive determinations of $V_{cb}$ and $V_{ub}$ in terms of right-handed currents is disfavored~\cite{Crivellin:2014zpa,Bernlochner:2014ova}. 

On the other hand, experiments have accumulated intriguing hints for the violation of LFU within recent years. In particular, the measurements of the ratios $R(D^{(*)})$~\cite{Lees:2012xj,Aaij:2017deq,Abdesselam:2019dgh} and $R(K^{(*)})$~\cite{Aaij:2017vbb,Aaij:2019wad} deviate from the SM expectation of LFU by more than $3\,\sigma$~\cite{Amhis:2019ckw,Murgui:2019czp,Shi:2019gxi,Blanke:2019qrx,Kumbhakar:2019avh} and $4\,\sigma$~\cite{Alguero:2019ptt,Aebischer:2019mlg,Ciuchini:2019usw,Arbey:2019duh}, respectively. In addition, the anomalous magnetic moments $(g-2)_\ell$ of charged leptons also measures the violation as they vanish in the massless limit. Here, there is the long-standing discrepancy in $(g-2)_\mu$ of about $3.7\,\sigma$~\cite{Bennett:2006fi,Aoyama:2020ynm,Aoyama:2012wk,Aoyama:2019ryr,Czarnecki:2002nt,Gnendiger:2013pva,Davier:2017zfy,Keshavarzi:2018mgv,Colangelo:2018mtw,Hoferichter:2019gzf,Davier:2019can,Keshavarzi:2019abf,Hoid:2020xjs,Kurz:2014wya,Melnikov:2003xd,Masjuan:2017tvw,Hoferichter:2018dmo,Hoferichter:2018kwz,Gerardin:2019vio,Colangelo:2017qdm,Colangelo:2017fiz,Bijnens:2019ghy,Colangelo:2019lpu,Colangelo:2019uex,Blum:2019ugy,Colangelo:2014qya} and a recently emerging deviation in the electron case of $2.5\,\sigma$, interestingly, with the opposite sign~\cite{Hanneke:2008tm,Parker:2018vye,Laporta:2017okg,Aoyama:2017uqe,Davoudiasl:2018fbb,Crivellin:2018qmi,Giudice:2012ms}.  Furthermore, it has been shown that LFU violating neutrino interactions with SM gauge bosons give an excellent fit of electroweak data, including LFU tests~\cite{Coutinho:2019aiy}.  
		
In this context, it seems natural to consider the discrepancies between the different determinations of $V_{ud}$ (or, equivalently, $V_{us}$ under the assumption of CKM unitarity), in particular the direct determination from $\beta$ decays, not as a sign of right-handed currents or as a violation of CKM unitarity, but rather as a sign of LFUV~\cite{Coutinho:2019aiy}. In fact, the most precise determination of $V_{ud}$ from $K\to \mu\nu$ involves muons, while $\beta$ decays can only have electrons in the final states. Therefore, we offer the novel perspective to use $\beta$ decays to search for LFUV and propose to study the observable
\beq
R(V_{us})=\frac{V_{us}^{K_{\mu2}}}{V_{us}^\beta}
\label{RVus}
\eeq
as the corresponding measure. As we will show below, this observable proves to be extremely sensitive to LFUV in the charged current (in particular in the muon sector) and even complements the picture described above, as it deviates by $2$--$3\,\sigma$ from unity. Including $K\to \pi\ell\nu$ and $V_{us}$ from $\tau$ decays into the analysis would even increase the tension towards the $4\,\sigma$ level~\cite{Belfatto:2019swo,Cheung:2020vqm,Grossman:2019bzp}.

{\it Beta decays and LFUV}.---We are interested in testing LFU of the charged current, i.e., of $W$--$\ell$--$\nu$ couplings, which we parameterize in terms of small corrections $\eps_{ij}$ according to
\beq
\Lagr \supset
-i\frac{{g_2}}{{\sqrt 2 }}{{\bar \ell }_i}{\gamma ^\mu }{P_L}{\nu _j}{W_\mu }\left( {{\delta _{ij}} + {\eps _{ij}}} \right),
\label{couplings}
\eeq
with the SM recovered for $\eps_{ij}\to 0$. Here we neglected tiny neutrino masses and set the leptonic mixing matrix to unity. Furthermore, we will disregard flavor-violating $\eps _{ij}$ parameters in the following since they are tightly bounded by radiative lepton decays $\ell\to \ell^\prime\gamma$ and lead to contributions that do not interfere with the SM in observables testing LFU. Note that in Eq.~\eqref{couplings} we simply parameterize the BSM effect by $\eps_{ij}$, but do not consider the $SU(2)_L$ gauge invariance in SMEFT as discussed in Ref.~\cite{Cirigliano:2009wk}.

The first crucial observation is that the corresponding modification of the charged current affects the determination of the Fermi constant $G_F$ from the muon life time~\cite{Tishchenko:2012ie}
\beq
\frac{1}{\tau_{\mu}}=\frac{(G_F^{\Lagr})^2m_{\mu}^5}{192\pi^3}(1+\Delta q)(1+\varepsilon_{ee}+\varepsilon_{\mu\mu})^2,
\eeq
where $G_F^{\Lagr}$ is the Fermi constant appearing in the Lagrangian (excluding BSM contamination) and $\Delta q$ subsumes the phase space, QED, and hadronic radiative corrections. Accordingly, we conclude that 
the Fermi constant measured in muon decay (extracted under the SM assumption) is related to the one at the Lagrangian level (containing the fundamental parameters $M_W$ and $g_2$) as
\beq
G_F=G_F^{\Lagr}(1+\eps_{ee}+\eps_{\mu\mu}).
\label{GFmod}
\eeq
This correction has to be taken into account whenever considering a weak decay, unless normalized to another branching ratio subject to the same correction.

This redefinition of the Fermi constant affects the determination of $V_{ud}$ from all $\beta$-decay observables in the same way
\beq
V_{ud}^\beta=V_{ud}^\Lagr\big(1-\eps_{\mu\mu}\big),
\eeq
again denoting by $V_{ij}^\Lagr$ CKM matrix elements without any BSM contamination, which therefore, by definition, fulfill CKM unitarity exactly. 
In particular, the indirect corrections introduced via $G_F$ imply that $\beta$-decay observables are actually sensitive to LFUV in the muon, not the electron sector. To construct the LFU violating observable in Eq.~\eqref{RVus} we further define
\beq
V_{us}^\beta\equiv\sqrt{1-\big(V_{ud}^\beta\big)^2-\big|V_{ub}\big|^2}
\simeq V_{us}^\Lagr\bigg[1+\bigg(\frac{V_{ud}^\Lagr}{V_{us}^\Lagr}\bigg)^2\eps_{\mu\mu}\bigg].
\eeq
It is this crucial enhancement by $(V_{ud}/V_{us})^2\sim 20$ that generates the amplified sensitivity to LFUV of our proposed observable $R(V_{us})$.

\begin{table}[t]
	\renewcommand{\arraystretch}{1.3}
	\centering
	\begin{tabular}{l c r}
		\toprule
		Observable & Measurement & Constraint\\\colrule
		$\frac{K\rightarrow\pi\mu\bar{\nu}}{K\rightarrow\pi e\bar{\nu}}\simeq 1+\eps_{\mu\mu}-\eps_{ee}$ & $1.0010(25)$~\cite{Moulson:Amherst} & $1.0(2.5)$\\
		$\frac{K\rightarrow\mu\nu}{K\rightarrow e\nu}\simeq 1+\eps_{\mu\mu}-\eps_{ee} $ &$0.9978(18)$~\cite{Ambrosino:2009aa,Lazzeroni:2012cx,Tanabashi:2018oca} & $-2.2(1.8)$\\
		$\frac{\pi\rightarrow\mu\nu}{\pi\rightarrow e\nu}\simeq 1+\eps_{\mu\mu}-\eps_{ee} $ & $1.0010(9)$~\cite{Aguilar-Arevalo:2015cdf,Czapek:1993kc,Britton:1992pg,Tanabashi:2018oca} & $1.0(9)$\\	
		$\frac{\tau\rightarrow\mu\nu\bar{\nu}}{\tau\rightarrow e\nu\bar{\nu}}\simeq 1+\eps_{\mu\mu}-\eps_{ee}$ & $1.0018(14)$~\cite{Amhis:2019ckw,Tanabashi:2018oca} & $1.8(1.4)$\\	
		$\frac{W\rightarrow\mu\bar{\nu}}{W\rightarrow e\bar{\nu}}\simeq 1+\eps_{\mu\mu}-\eps_{ee}$ & $0.9960(100)$~\cite{Pich:2013lsa,Schael:2013ita} & $-4(10)$\\	
		$\frac{B\rightarrow D^{(*)}\mu\nu}{B\rightarrow D^{(*)}e\nu}\simeq 1+\eps_{\mu\mu}-\eps_{ee}$ & $0.9890(120)$~\cite{Jung:2018lfu} & $-11(12)$\\
		\hline
		$R(V_{us})\simeq 1-\big(\frac{V_{ud}}{V_{us}}\big)^2\eps_{\mu\mu}$ & 
		 $0.9891(33)$~\cite{Seng:2018yzq} & $0.58(17)$\\
		 & $0.9927(39)$~\cite{Czarnecki:2019mwq} & $0.39(21)$\\
		\botrule
	\end{tabular}
	\caption{Ratios sensitive to LFUV in the $\mu$--$e$ sector, indicating the dependence on the LFU violating parameters $\eps_{ij}$. For $R(V_{us})$ we give the values corresponding to the radiative corrections from Refs.~\cite{Seng:2018yzq,Czarnecki:2019mwq}. The last column gives the constraints on $\big(\eps_{\mu\mu}-\eps_{ee}\big)\times 10^3$ and $\eps_{\mu\mu}\times 10^3$, respectively.}  \label{tab:LFUtest}
\end{table}

Before turning to the numerical analysis, we compare the sensitivity of $R(V_{us})$ to that of other probes of LFUV.
 Apart from $K_{\ell 3}$ decays, as given in Eq.~\eqref{RKl3} below, this includes $K_{\ell 2}$ and $\pi_{\ell 2}$, $\tau\to\ell \nu\bar \nu$, and $W\to \ell\nu$, see Table~\ref{tab:LFUtest} for their dependence on the $\eps_{ii}$ as well as current experimental constraints. Concerning $B$ decays, only $B\to D^{(*)}e\nu/B\to D^{(*)}\mu\nu$ provides a relevant constraint~\cite{Jung:2018lfu}.

\begin{figure}[t]
	\centering
	\includegraphics[width=0.8
	\linewidth]{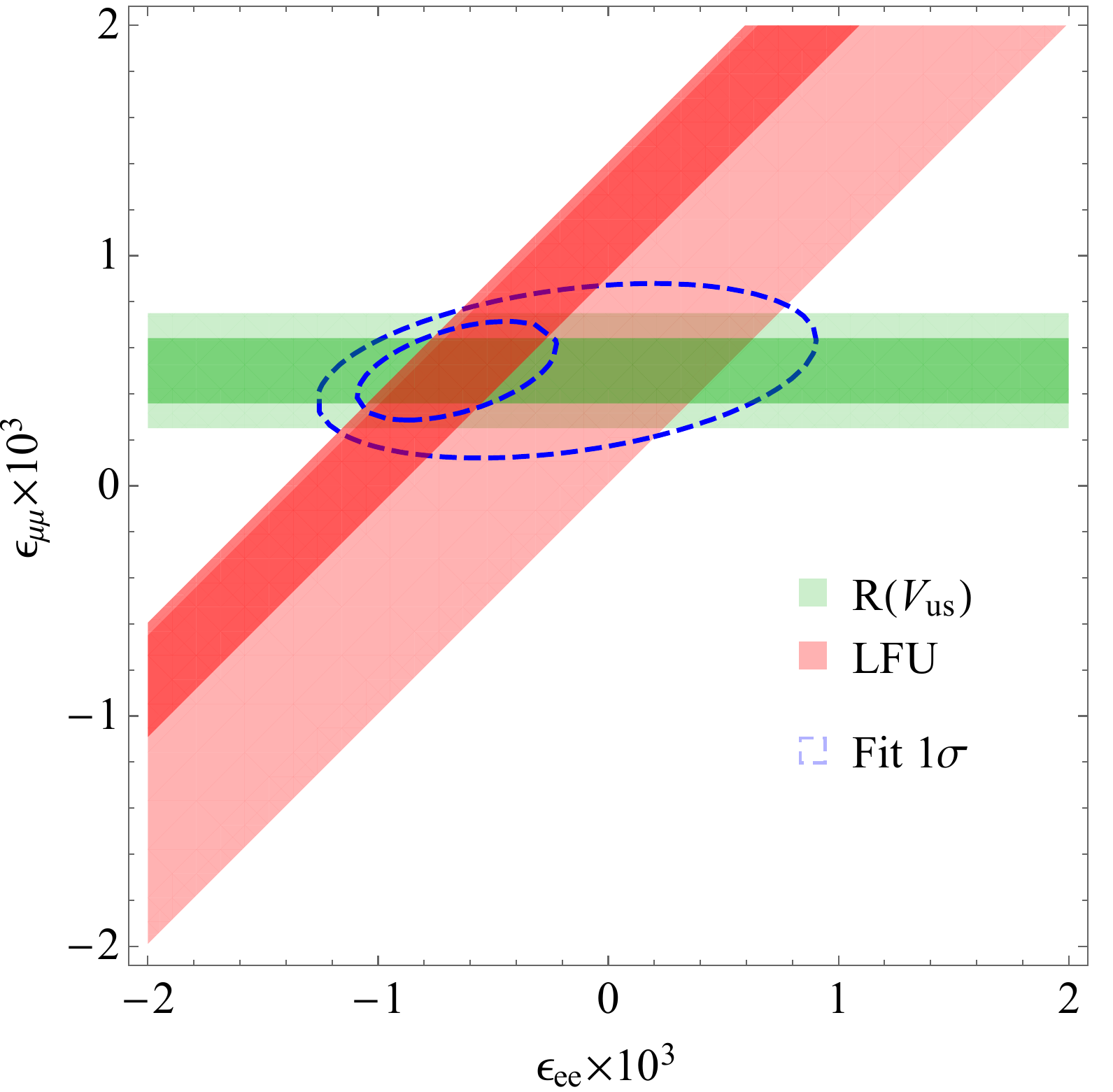}
	\caption{Fit to modified $W$--$\ell$--$\nu$ couplings. Here the light bands (large ellipse) correspond to the current status ($1\,\sigma$), where the band for $R(V_{us})$ has been increased to include both the CMS and the SGPR results. The dark bands (small ellipse) correspond to a future scenario in which the errors on $\tau\rightarrow\mu\nu\bar{\nu}/\tau\rightarrow e\nu\bar{\nu}$ and $\pi\rightarrow\mu\nu/\pi\rightarrow e\nu$ have decreased by a factor 3 due to forthcoming Belle II  and PEN/PiENu results, respectively. The projected $R(V_{us})$ band 
		assumes that the difference between the radiative corrections of CMS and SGPR has been understood and that a competitive $V_{ud}$ determination from neutron decay has become available.}
	\label{fig:Fit}
\end{figure}

A crucial advantage of $R(V_{us})$ is that all these ratios are sensitive to the difference $\eps_{\mu\mu}-\eps_{ee}$, not LFUV in either sector separately, and thus can only test LFU in case the $\eps_{ii}$ differ. In addition, none of the other ratios can probe LFU at a level below $\Order(10^{-3})$ yet, demonstrating the superior sensitivity of $R(V_{us})$ thanks to the CKM enhancement. We illustrate the comparison and complementarity of $R(V_{us})$ with respect to the other observables testing LFU in the $W$--$\ell$--$\nu$ couplings in Fig.~\ref{fig:Fit}, anticipating the results from the following numerical analysis. 

We start from the master formula for superallowed $\beta$ decays~\cite{Hardy:2018zsb}
\beq
|V_{ud}|^2=\frac{2984.432(3)\,\text{s}}{\Ft(1+\Delta_R^V)},
\eeq
with $\Ft$-value $\Ft=3072.07(63)\,\text{s}$~\cite{Hardy:2018zsb} and two sets of radiative corrections
\begin{align}
\Delta_R^V\big|_\text{SGPR}&=0.02467(22)\qquad \text{\cite{Seng:2018yzq}},\\
 \Delta_R^V\big|_\text{CMS}&=0.02426(32)\qquad \text{\cite{Czarnecki:2019mwq}},
\end{align}
leading to
\begin{align}
V_{ud}^\beta\big|_\text{SGPR}&=0.97370(14), &
 V_{us}^\beta|_{\text{SGPR}} &=0.22782(62),\notag\\
 V_{ud}^\beta\big|_\text{CMS}&=0.97389(18),&
 V_{us}^\beta\big|_{\text{CMS}}&=0.22699(78),
 \label{Vusbeta}
\end{align}
where we used $|V_{ub}|=0.003683$ from Refs.~\cite{CKMfitter:2019,Charles:2004jd}, although the precise value of $|V_{ub}|$ is immaterial here.
In addition to the universal electroweak corrections $\Delta_R^V$, it has been pointed out in Refs.~\cite{Seng:2018qru,Gorchtein:2018fxl} that also $\Ft$ may be subject to additional nuclear corrections. The final recommendation $\Ft=3072(2)\,\text{s}$ in Ref.~\cite{Gorchtein:2018fxl} leaves the central value largely unchanged, but implies a significant increase in uncertainty. Since the role of nuclear corrections in $0^+\to 0^+$ transitions is far from settled, we continue to employ $\Ft$ from Ref.~\cite{Hardy:2018zsb}, keeping in mind that the nuclear uncertainties are potentially underestimated.

For the neutron life time we use the master formula~\cite{Czarnecki:2018okw,Czarnecki:2019mwq} 
\beq
|V_{ud}|^2\tau_n(1+3g_A^2)(1+\Delta_\text{RC})=5100.1(7)\,\text{s},
\eeq
with radiative corrections
\beq
\Delta_\text{RC}^\text{SGPR}=0.03992(22),\qquad
\Delta_\text{RC}^\text{CMS}=0.03947(32).
\eeq
The experimental value for $g_A$ from the asymmetry parameter is confronted with
the lattice-QCD calculation $g_A=1.271(13)$~\cite{Chang:2018uxx} (see Ref.~\cite{Gupta:2018qil} for a critical assessment of the error estimate).  Finally, pion $\beta$ decay gives~\cite{Pocanic:2003pf,Czarnecki:2019iwz}
\beq
V_{ud}^\beta=0.9739(29).
\eeq

With the enhanced sensitivity to LFUV originating solely from $V_{us}^\beta$, it is less crucial which determination enters the numerator in Eq.~\eqref{RVus}. $V_{us}$ can be determined directly from semi-leptonic kaon decays $K_{\ell 3}$. Using the compilation from Ref.~\cite{Moulson:Amherst} (updating Ref.~\cite{Moulson:2017ive}) as well as the form factor normalization $f_+(0) = 0.9698(17)$~\cite{Carrasco:2016kpy,Bazavov:2018kjg,Moulson:Amherst}, we have
\begin{align}
 V_{us}^{K_{\mu 3}}&=0.22345(54)(39)=0.22345(67),\notag\\
V_{us}^{K_{e 3}}&=0.22320(46)(39)=0.22320(61),
\label{VusKl3}
\end{align}
where the first error refers to experiment and the second to the form factor.
LFUV affects these values according to
\begin{align}
V_{us}^{K_{\mu 3}}&=V^{\Lagr}_{us}\left(1-\eps_{ee}\right),\notag\\
V_{us}^{K_{e 3}}&=V^{\Lagr}_{us}\left(1-\eps_{\mu\mu}\right),
\end{align}
leading to the constraint 
\beq
R(K_{\ell3})=\frac{V_{us}^{K_{\mu 3}}}{V_{us}^{K_{e 3}}}
=1+\eps_{\mu\mu}-\eps_{ee}=1.0010(25),
\label{RKl3}
\eeq
where several uncertainties cancel in the ratio~\cite{Moulson:Amherst}.

For the purely leptonic kaon decays $K_{\ell2}$ one typically considers the ratio $K\!\to\!\mu\nu$ over $\pi\!\to\!\mu\nu$ to cancel the dependence on absolute decay constants. This allows one to directly determine 
 $V_{us}^\Lagr/V_{ud}^\Lagr$ once the ratio of decay constants $f_{K^\pm}/f_{\pi^{\pm}}$ as well as the treatment of isospin-breaking corrections are specified~\cite{Cirigliano:2011tm,DiCarlo:2019thl}. Here, we follow the strategy in Ref.~\cite{Moulson:Amherst} to use the recent results from lattice QCD~\cite{DiCarlo:2019thl}, at the same time adjusting the FLAG average~\cite{Aoki:2019cca} back to the isospin limit 
$f_{K^\pm}/f_{\pi^{\pm}}=1.1967(18)$~\cite{Dowdall:2013rya,Carrasco:2014poa,Bazavov:2017lyh}, to obtain 
\beq
V_{ud}^{K_{\mu2}}=0.97427(10),\qquad V_{us}^{K_{\mu 2}}=0.22534(42).
 \label{VusKl2}
  \eeq
The tension with the determinations from $K_{\ell 3}$~\eqref{VusKl3} cannot be explained with LFUV. For definiteness, we will use the $K_{\mu 2}$ value as reference point in Eq.~\eqref{RVus}, given that in contrast to $K_{\ell 3}$ it is, by definition, not sensitive to LFUV. Note that this is a conservative choice in the sense that the value of $R(V_{us})$ lies closer to unity if $K_{\mu 2}$ and not $K_{\ell 3}$ is used. One could extend the analysis further to determinations from $\tau$ decays, see Ref.~\cite{Amhis:2019ckw}, but here the errors are larger and again there are tensions between the inclusive and exclusive determinations. Therefore, for simplicity we restrict the analysis to observables sensitive only to the $\mu$--$e$ sector.

Numerically, $V_{us}^\beta$ from Eq.~\eqref{Vusbeta} and $V_{us}^{K_{\mu 2}}$ from Eq.~\eqref{VusKl2} provide the constraint
\beq
\eps_{\mu\mu}\big|_\text{SGPR}=0.00058(17),\qquad
\eps_{\mu\mu}\big|_\text{CMS}=0.00039(21),
\label{RVus_epsmumu}
\eeq
and thus a sensitivity to LFUV below $\Order(10^{-3})$. If instead the $V_{us}$ values from $K_{\ell 3}$ decays were used, the central values would increase to $\eps_{\mu\mu}\sim 0.001$ with similar errors as in Eq.~\eqref{RVus_epsmumu}, thus implying a much higher significance.

{\it Future Prospects}.---Future improvements of the analysis presented here are foreseen at several frontiers:
(1) the numerical value and accuracy of $R(V_{us})$ could be consolidated with improved radiative corrections;
(2) improved experimental input for neutron life time and asymmetry parameter could make the resulting $V_{ud}$ determination competitive with superallowed $\beta$ decays;
(3) new data could shed light on the tension between $K_{\ell2}$ and $K_{\ell 3}$ decays.

First, for the superallowed $0^+\to 0^+$ transitions the main uncertainty at this point originates from radiative corrections, both universal radiative corrections that also affect neutron decay as well as additional nuclear effects. The latter should be amenable to refined calculations with modern nuclear structure theory (see, e.g., Ref.~\cite{Gysbers:2019uyb} for a recent ab-initio calculation of nuclear $\beta$ decays and Ref.~\cite{Cirgiliano:2019nyn} for a discussion of the nuclear theory requirements). Meanwhile, improving the universal radiative corrections rests on a better understanding of the nucleon matrix elements $\langle p|T\{j^\mu_\text{em}j^\nu_{w,A}\}|n\rangle$ involving the electromagnetic current $j^\mu_\text{em}$ and the axial part of the charged weak current $j^\mu_{w,A}$, for which either new input from experiment or lattice QCD~\cite{Seng:2019plg} is required.

Second, the measurement of $\lambda$ is currently dominated by the measurement of Ref.~\cite{Markisch:2018ndu} (with small changes if earlier results from Refs.~\cite{Mund:2012fq,Brown:2017mhw} are included in the average), leading to a relative precision of $4\times 10^{-4}$. There are several ongoing and planned developments that promise to extend the sensitivity towards or even beyond the level of $10^{-4}$~\cite{Fry:2018kvq,Soldner:2018ycf,Wang:2019pts,Serebrov:2019puy,Markisch:Amherst}. 
 To establish the $V_{ud}$ determination from neutron decay at a level competitive with superallowed $\beta$ decays commensurate improvements in the life time are necessary. In addition to the long-standing discrepancy between bottle and beam measurements (see Refs.~\cite{Czarnecki:2018okw,Wietfeldt:2011suo} for reviews), also the difference between recent bottle measurements~\cite{Pattie:2017vsj,Ezhov:2014tna,Serebrov:2017bzo} currently leads to a non-negligible scale factor in the PDG average~\cite{Tanabashi:2018oca}. Fortunately, there are plans to probe $\tau_n$ at a level down to hundreds of ms~\cite{Gaisbauer:2016kan,Ezhov:2018cta,Callahan:2018iud,Saunders:PSI}. 

Third, preliminary data on $K_{\ell 3}$ decays exist from the OKA~\cite{Yushchenko:2017fzv} and KLOE-2~\cite{Babusci:2019gqu} experiments, with further input potentially from LHCb~\cite{Junior:2018odx}, NA62, and TREK~\cite{Moulson:Amherst}. Further insights on $V_{us}$ could be obtained from semi-leptonic hyperon decays~\cite{Cabibbo:2003ea,Cabibbo:2003cu,FloresMendieta:2004sk,Mateu:2005wi,Geng:2009ik} given renewed experimental interest at BESIII~\cite{Li:2016tlt,Ablikim:2019hff}, but would also require progress in lattice-QCD calculations of the hyperon form factors~\cite{Sasaki:2017jue}.
All of these developments (1)--(3) should help establish or refute the current $2$--$3\,\sigma$ hint~\eqref{RVus_epsmumu} for LFUV in $\beta$ decays.

In addition, there are several experimental developments dedicated to improving the LFU tests in Table~\ref{tab:LFUtest}. The J-PARC E36 experiment aims at improving
$K\to\mu\nu/K\to e\nu$~\cite{Shimizu:2018jgs}, while the ratio of $K_{\ell 3}$ decays could profit from the developments mentioned above. A similar sensitivity as in $R(V_{us})$ may be possible for $\tau\to \mu\nu\bar\nu/\tau\to  e\nu\bar\nu$  at Belle II~\cite{Kou:2018nap}, where approximately one order of magnitude more $\tau$ leptons will be produced than at Belle or BaBar. At this level, one would directly probe the same parameter space as in Eq.~\eqref{RVus_epsmumu}, barring of course a significant cancellation between $\eps_{\mu\mu}$ and $\eps_{ee}$. The most promising observable, however, is currently $\pi\to\mu\nu/\pi\to e\nu$, for which the PEN~\cite{Glaser:2018aat} and PiENu~\cite{Mischke:2018qmv} experiments anticipate in the near future an improvement by more than a factor $3$, which would bring the limit on $\eps_{\mu\mu}-\eps_{ee}$ well below $\Order(10^{-3})$ as well.
Taking into account all these potential improvements, we also included an optimistic but realistic projection of future constraints 
in Fig.~\ref{fig:Fit}.

Moving beyond pure modification of $W$--$\ell$--$\nu$ couplings, one can see from an analysis of gauge-invariant dimension-6 operators that a simultaneous modification of $Z$--$\ell$--$\ell$ and/or $Z$--$\nu$--$\nu$ is unavoidable: there are only two operators modifying these couplings~\cite{Buchmuller:1985jz,Grzadkowski:2010es}, so that the effects in at most one of these three couplings can be canceled. The LEP bounds on $Z$--$\ell$--$\ell$ couplings are already now at the per mille level~\cite{Schael:2013ita} and also the bounds on the invisible $Z$ width  (corresponding to $Z$--$\nu$--$\nu$ in the SM) are excellent. These bounds could be significantly improved by future $e^+e^-$ colliders such as an ILC~\cite{Baer:2013cma}, CLIC~\cite{deBlas:2018mhx}, or an FCC-ee~\cite{Abada:2019lih,Abada:2019zxq}. Furthermore, $W$ pair production will allow for a direct determination of $W\to\mu\nu/W\to e\nu$. In particular, an FCC-ee could produce up to $10^8$ $W$ bosons (compared to the LEP number of $4\times 10^4$), leading to an increase in precision that would render a direct discovery of LFUV in $W$--$\ell$--$\nu$ conceivable. 

{\it Conclusions}.---$\beta$ decays are high-precision low-energy tests of the SM. While so far these decays were used to constrain CKM unitarity or right-handed currents (from superallowed $\beta$ decays and neutron decay), we showed in this Letter that they are also an exquisite probe of LFU: due to the conventional definition of the Fermi constant they actually probe LFU in the muon sector, with a sensitivity CKM-enhanced  by $(V_{ud}/V_{us})^2\sim 20$. Therefore, we proposed and examined the ratio~\eqref{RVus}
to test LFU. This measure $R(V_{us})$ is complementary to other probes of LFU, most notably $\pi\to\mu\nu/\pi\to e\nu$ and $\tau\to \mu\nu\bar\nu/\tau\to  e\nu\bar\nu$, which are
sensitive to the difference of muon and electron couplings.
  
Current data show a deviation of $R(V_{us})$ from unity at the level of $2$--$3\,\sigma$, depending on assumptions for the radiative corrections. In light of the accumulated hints for LFUV in $R(K^{(*)})$, $R(D^{(*)})$, and $(g-2)_{\mu,e}$, it seems natural to consider $\beta$ decays as a probe of LFU and refine complementary tests of LFU with this connection in mind. We discussed several avenues how the present constraints can be improved in the future, including experimental developments in kaon and neutron decays, which, however,  
should be accompanied by adequate efforts on the theory side aiming at improving our understanding of radiative corrections. 
Similar improvements are anticipated in related tests of LFU, with results expected soon from the PEN and PiENu experiments, while 
future $e^+e^-$ colliders, in particular the FCC-ee, would even have the potential to directly observe LFUV in $W$ decays.

\begin{acknowledgments}
 We thank CERN for support during the Theory Institute ``New Physics on the Low-Energy Precision Frontier,'' where part of this work was performed. 
Support by the Swiss National Science Foundation, under Project Nos.\ PP00P21\_76884 (A.C.) and PCEFP2\_181117 (M.H.), and by the DOE (Grant No.\ DE-FG02-00ER41132) is gratefully acknowledged.
\end{acknowledgments}

\end{document}